\documentclass[a4paper,10pt,prl,twocolumn,showpacs,aps,floats,superscriptaddress]{revtex4}

\usepackage{epsfig}

\begin{document}

\title{Critical load and congestion instabilities in scale-free networks}
\author{Yamir Moreno}
\affiliation{The Abdus Salam International Centre for Theoretical Physics, 
P.O. Box 586, 34100 Trieste, Italy}
\author{Romualdo Pastor-Satorras}
\affiliation{ Departament de
F\'{\i}sica i Enginyeria Nuclear, Universitat Polit\`{e}cnica de
Catalunya, Campus Nord, M\`{o}dul B4, 08034 Barcelona, Spain}
\author{Alexei V{\'a}zquez}
\affiliation{International School for Advanced Studies SISSA/ISAS,
  via Beirut 4, 34014 Trieste, Italy}
\author{Alessandro Vespignani}
\affiliation{The Abdus Salam International Centre for Theoretical Physics, 
P.O. Box 586, 34100 Trieste, Italy}
\date{\today}

\widetext
\begin{abstract}
We study the tolerance to congestion failures in communication
networks with scale-free topology. The traffic load carried by each
damaged element in the network must be partly or totally redistributed
among the remaining elements. Overloaded elements might fail on their
turn, triggering the occurrence of failure cascades able to isolate
large parts of the network.  We find a critical traffic load above
which the probability of massive traffic congestions destroying the
network communication capabilities is finite.
\end{abstract}


\maketitle

Complex heterogeneous connectivity patterns have been recently
identified in several natural and technological networks
\cite{strogatz,bara02,doro}.  The Internet and the World-Wide-Web
(WWW) networks, where nodes represent routers or web pages and edges
physical connections or hyper-links, appear to have a topology
characterized by the presence of ``hubs'' with many connections to
peripherical nodes. Empirical evidence recently collected shows that
this distinctive feature finds its statistical characterization in the
presence of heavy-tailed degree distributions
\cite{bara00,fal,broder00,gov00,gcalda,gov01,caida,romu01}.  In the
Internet, for instance, the statistical analysis reveals that the
degree distribution $P(k)$, defined as the probability that any node
has $k$ links to other nodes, is well approximated by a power-law
behavior $P(k)\sim k^{-\gamma}$, with $\gamma\approx 2.2$ \cite{fal,gov00,romu01}.
This makes the Internet a capital example of the recently identified
class of scale-free (SF) networks \cite{bara02}.  The statistical
physics approach has been proved to be a valuable tool for the study
of complex networks, and several interesting results concerning
dynamical processes taking place on complex networks have been
recently reported. In particular, the absence of the percolation
\cite{newman00,havlin01} and epidemic \cite{pv01a,virusreview}
thresholds in SF networks has a large impact because of its potential
practical implications.  The absence of the percolating threshold,
indeed, prompts to an exceptional tolerance to random damages
\cite{barabasi00}. This is a property that assumes a great importance
in communication networks, guaranteeing the connectivity capabilities
of the system.

Percolation properties of SF networks refer only to the static
topological connectivity properties \cite{newman00,havlin01}. On the
other hand, in the Internet and other communications networks, many
instabilities are due to traffic load congestions \cite{labo1,labo2}.
The traffic load carried on the failing nodes or connections is
automatically diverted to alternative paths on the networks and
instabilities can spread from node to node by an avalanche of traffic
congestions and overloads. For instance, route flaps have led to the
transient loss of connectivity for large portions of the Internet.
These instabilities are thus of a dynamical nature and depend on how
information is routed and distributed in the network. The models
proposed so far, however, deal with regular structures
\cite{jap1,jap2,watts02} and do not take into account the complex topology of
SF networks.

In this paper, we propose a simple model aimed at the study of failure
cascades generated by the redistribution of traffic load by congested
links or nodes in SF networks.  We find that the system behavior
depends on the average traffic load imposed to the network. Above a
critical value of the average traffic load, a single failure has a
finite probability of triggering a congestion avalanche affecting a
macroscopic part of the network. The present analysis thus reveals the
existence of a transition from a free phase to a congested one as a
function of the amount of traffic carried by the network. Contrary to
what happens for the static percolation transition
\cite{havlin01,newman00}, loaded SF networks exhibit a finite
threshold above which the system can develop macroscopic instabilities
with respect to small damages if we consider the dynamics of the
traffic carried on top of them. The results provided here represent a
first step towards a more complete modeling of traffic instabilities
in real communication networks.

In order to include the degree fluctuations of SF networks we shall
use in the following the Barab\'{a}si-Albert model \cite{bar99}. This
is a stochastic growth model in which one starts from a small number
$m_0$ of nodes and at each time step a new node is introduced. The new
node is connected preferentially to $m$ old ones (for the simulations
we used $m=3$) with a probability $\Pi(k_i)=k_i/\sum_j k_j$ proportional
to the degrees $k_i$ of the nodes.  The repeated iteration of this
scheme gives as a result a complex network with a topological
structure characterized by a power-law degree distribution
$P(k)=2m^2k^{-\gamma}$ with $\gamma=3$ and average degree $\langle k \rangle=2m$. In
principle, one might also consider a more general class of complex
networks with variable power-law degree distributions
\cite{bara02,doro}.

To simulate the flow of data packets on SF networks, taking into
account the load redistribution in case of damages, we need to specify
the initial state of the network; i.e. the load of traffic flowing
through each link. An estimate of such load, assuming that the routing
takes place following the minimum path, is given by the total number
of shortest paths between any two nodes in the network that pass
through the node $i$. This magnitude is called betweenness or load
\cite{new01,goh01} and has been recently studied in SF networks. This
property of the network can also be defined in terms of links.  In
real systems, however, the amount of traffic carried by each link is a
fluctuating quantity that depends on many variables such as number of
users, routing agreements, and available bandwidth.  For this reason,
we associate to each link connecting the nodes $i$ and $j$ of the
network a load $\ell_{i,j}$ drawn from a probability distribution that
specifies the initial traffic load of the system.  For simplicity, we
have considered a uniform distribution $U(\ell)$ for $0< \ell < 1$, taking
the form
\begin{displaymath}
  U(\ell) = \left\{ 
    \begin{array}{cll}
      \frac{1}{2 \langle \ell \rangle }, & \ell \in [0, 2 \langle \ell \rangle ] &
      \mathrm{if} \, \langle \ell \rangle \leq  0.5\\ &&\\
      \frac{1}{2(1-  \langle \ell \rangle) }, & \ell \in [2 \langle \ell \rangle -1 ,1
      ] & \mathrm{if} \,\langle \ell \rangle \geq 0.5
    \end{array} \right..
\end{displaymath}
This uniform distribution implies that the minimum initial load
carried by a link is bounded by a nonzero value for an average load
$\langle l\rangle > 0.5$, which means that there will be no links with load
smaller than this lower bound. The results reported in this paper were
obtained with the initial distribution $U(\ell)$. In
order to test the universality of the critical behavior, we have also
considered the distribution $F(\ell)= (\langle \ell \rangle^{-1} -1) (1-\ell)^{(\langle \ell
  \rangle^{-1} -2)}$, $\ell \in [0, 1 ]$, which allows the existence of links
with a very small load, irrespective of the average load $\langle\ell\rangle$
flowing through the system.  Both distributions $U(\ell)$ and $F(\ell)$
yield the same qualitative results. Along with the load, we associate
to each link the same capacity $C$ that, without loss of generality,
we fix equal to one.  This choice can be considered as a first
approximation since the actual difference between line bandwidths in
communication networks can be large.  In this perspective, we consider
as the most important source of heterogeneity the flow of different
amounts of load through the network.
 
The dynamics of the model is defined by a simple threshold process.  A
link is selected at random and overloaded by raising its traffic.
When the load carried by a link is $\ell_{i,j}>C$, i.e. when it exceeds
the link's capacity, the link is considered congested and the load it
carries is diverted among its (not overloaded) neighboring links. This
amounts to consider that the time scale of the local congestion is
greater than the time scale characterizing the reorganization of the
routing procedure. The redistribution of the load on its turn might
provoke that other links become overloaded, thus triggering a cascade
of failures.  We have explored two physically different settings of
the load redistribution rule.  The first consists on equally
distributing the load of a congested link among the non-congested
neighboring links. We refer to it as the deterministic redistribution
rule, respectively. The second case will be called random
redistribution because when a link is overloaded, a random amount of
load is redistributed to each of the remaining working links in its
neighborhood.  Finally we note that in the rare event in which the
congested link has no active neighbors, its load can be equally shared
among all the remaining working lines of the network or just be
considered as lost from the network. This amounts to a conserved or
dissipative redistribution rule.  Many physical systems display
criticality only when energy is conserved \cite{jen98,marro}. In
distributed networks such as the Internet, however, it is common to
discard packets if there is not a route available at the moment. As we
shall see in the following, the results do not depend qualitatively on
the conserved nature of the traffic load.

\begin{figure}[t]
\begin{center}  
\epsfig{file=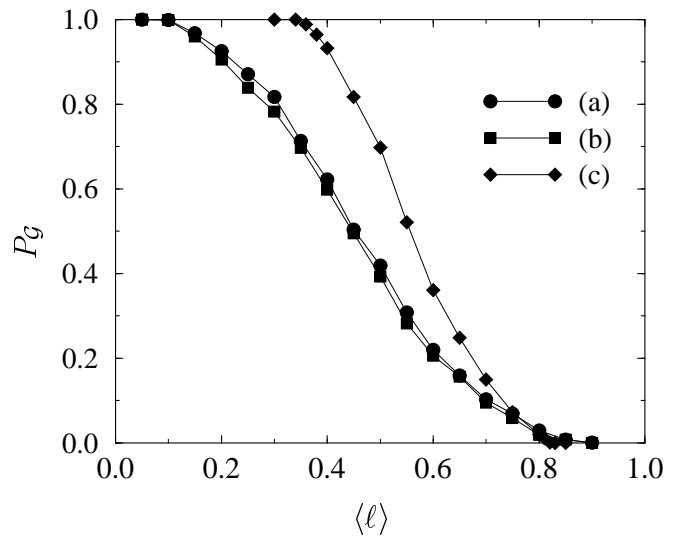,width=3.4in}
\end{center}
\caption{Phase diagram of the system. $P_\mathcal{G}$ measures the
  probability of having a giant component of active nodes
  $\mathcal{G}$ of the order of the system size. Three different
  definitions of the model are represented: (a) random and
  dissipative, (b) random and conserved, and (c) deterministic and
  conserved. In all cases the size of the system has been set to
  $N=10^4$ nodes ($3 \times 10^4$ links).}
\label{figure1}
\end{figure}

We have performed large-scale numerical simulations by applying
repeatedly the rules stated above on BA networks. The sizes of the
networks used in the simulations range from $N=5 \times 10^3$ nodes ($15 \times
10^3$ links) to $N=10^5$ nodes ($3 \times 10^5$ links). All numerical
results have been obtained by averaging over 10 different networks
and, at least, 100 different realizations of the initial load
distribution.

In order to inspect the occurrence of dynamic instabilities, we
construct the phase diagram of the system. The order parameter can be
identified as the probability $P_\mathcal{G}$ of having a giant
component $\mathcal{G}$ of connected nodes with size of the order of
the network size.  The giant component is defined as the largest
component of the network made by nodes connected by active links,
after the system has reached a stable state (when $\ell_{i,j} < C$ for
all $i$ and $j$). The existence of a giant component implies that a
macroscopic part of the network is still functional.  If the giant
component of the network is zero, the communication capabilities of
the network are destroyed and a congestion of the order of the system
size builds up. It is worth noticing that although we have determined
the giant component size in terms of nodes and the dynamical rules of
the model are expressed in terms of links, the results are completely
equivalent since a connected node is defined as a node with at least
one active link.

\begin{figure}[t]
\begin{center}
\epsfig{file=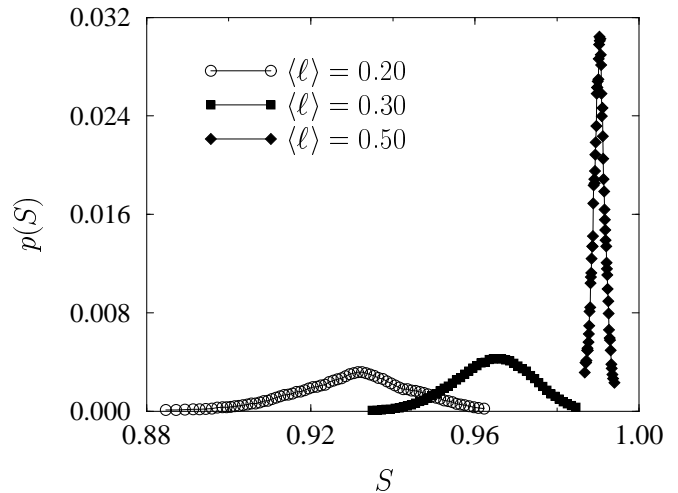,width=3.4in}
\end{center}
\caption{ Distribution of the isolated network size  $p(S)$ as a
  function of $S$ for several values of $\langle \ell \rangle$. The network is
  formed by $N=10^4$ nodes.  }
\label{figure2}
\end{figure}

In Fig.~\ref{figure1} we plot the order parameter $P_\mathcal{G}$ as a
function of the average load $\langle\ell\rangle$.  At low values of the average
load, the network always reaches a stable state in which the number of
isolated nodes is very small and with probability $P_\mathcal{G}=1$
the network has a giant component of connected nodes of the order of
the system size.  When increasing the load imposed on the network, the
system starts to develop instabilities. In particular, above a
critical load $\langle\ell\rangle_c^I$, whose value depends on the model
considered, with a finite probability the system evolves to a
congested state without giant component of connected nodes; i.e. the
largest set of connected active nodes has a density of order $N^{-1}$.
This implies a probability of having a giant component
$P_\mathcal{G}<1$, which is decreasing as the load is progressively
increased.  At an average load $\langle\ell\rangle_c^{II}\simeq 0.82$ we get that
$P_\mathcal{G}=0$, signalling that, with probability one, any
instability will propagate until the complete fragmentation of the
network.  It is worth remarking that this scenario is rather different
from the percolation one in which the probability of having a giant
component is abruptly dropping from one to zero at the transition
point. Here, the probability decays continuously to zero and we have a
wide region of $\langle\ell\rangle$ where the initial instability can trigger a
destructive congestion with probability $1-P_\mathcal{G}$.
Fig.~\ref{figure2} illustrates the probability $p(S)$ that the
isolated network has a size $S$ in the case that no giant component of
connected nodes has survived. The distribution is rather peaked also at
relatively small values of $\langle \ell \rangle$, almost affecting the totality
of the network.

\begin{figure}[t]
\begin{center}
\epsfig{file=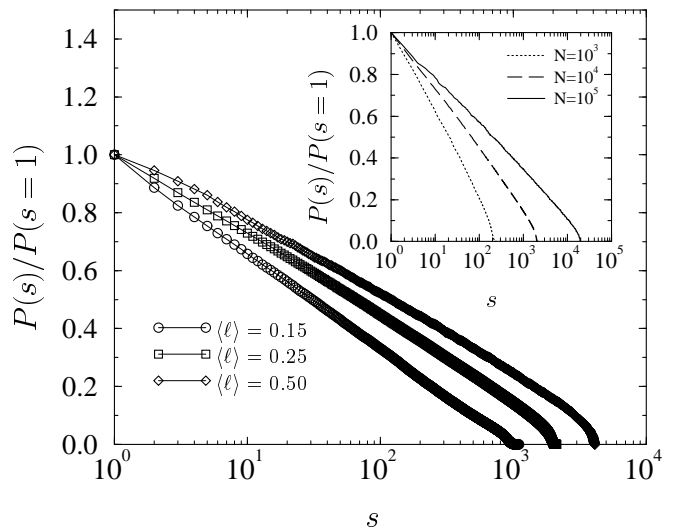,width=3.4in}
\end{center}
\caption{Cumulative distributions of avalanche sizes for different
  values of the average load handled by the network for the random and
  dissipative definition of the model.  The straight line in a
  linear-log plot indicates that the probability avalanche
  distribution follows a power law with exponent $-1$. The inset shows
  the scaling of the cumulative size of congested lines as the network
  grows in size for $\langle \ell \rangle=0.25$.}
\label{figure3}
\end{figure}

The phase diagram obtained in Fig.~\ref{figure1} points out that the
value of the average load at which $P_\mathcal{G}=0$ is relatively
high. On the other hand, the value at which $P_\mathcal{G}$ is
appreciably smaller than one is well below the theoretical capacity of
the network measured as the capacity $C=1$ of the individual links
($\langle\ell\rangle_c^I\simeq 0.15 - 0.3$, see Fig.~\ref{figure1}).  This evidence
defines a wide region of load values in which there is a small but
finite probability that a small instability propagates through the
system and could explain why in real communication networks
congestions of diverse degree can be encountered from time to time.
In Fig.\ \ref{figure1} we also report curves obtained for the
different dynamical rules defined previously and it is interesting to
remark that the highest level of stability has been obtained for the
deterministic and conservative version of the model ($\langle\ell\rangle_c^I\simeq
0.3$). Moreover, the dynamics of the transition does not depend on the
distribution of initial loads and the network size $N$.

Another way to shed light on the congestion dynamics is to inspect the
process of generation of the congestion instability.  We define the
size $s$ of a congestion burst or avalanche as the total number of
simultaneously overloaded links.  The cumulative distribution $P(s)$
of avalanches of size larger than $s$ for several values of the
average load imposed on the network and four different system sizes
have been plotted in Fig.~\ref{figure3} for the random and dissipative
definition of the present model.  The main plot is in log-linear
scale, so that a straight line corresponds to a power law of the form
$p(s) \sim s^{-1}$ for the probability of observing an avalanche of size
$s$. Power-laws with exponent $-1$ have been found for several
characteristic features of Internet traffic such as latency times,
queue lengths, and congestion lengths \cite{jap1,mag,will}. In the
figure we focus on region close to the stable region $\langle \ell \rangle\simeq
0.20$, which means that the power-law behavior extends to values far
from the instability transition. This fact confirms that it is not
necessary that the network operates very close to a critical point in
order to observe power-laws in the distribution of several quantities.
The inset in Fig.~\ref{figure3} shows that the cumulative size of
overloaded links also scales with the system size, the scaling
dynamics, however, remaining the same. This may help understand why
power-law distributions observed in real communication networks have
been measured for different network sizes, i.e. both for local
networks and for networks that extend to a very large scale.

In summary, we have introduced a simple threshold model aimed at the
description of instabilities due to load congestion that takes into
account the topological properties of SF networks. The results
obtained point out that the network can freely handle traffic up to
some critical average load $\langle\ell\rangle_c^I$.  Above this level the network
faces partial congestions that start to build up local bottlenecks in
various places and small instabilities might trigger macroscopic
outages with a finite probability.  Above a critical load value
$\langle\ell\rangle_c^{II}\simeq 0.82$ any small instability leads to the whole
network collapse.  In the intermediate region of network load, the
number of simultaneous line casualties follows a power-law resembling
what has been observed in experimental studies of the Internet.  We
hope that our work will provide hints for accurate modeling of the
Internet and the WWW large-scale traffic behavior.

\begin{acknowledgments}
This work has been partially supported by the European Commission -
Fet Open project COSIN IST-2001-33555. R.P.-S. acknowledges financial
support from the Ministerio de Ciencia y Tecnolog\'{\i}a (Spain).
\end{acknowledgments}

\end{document}